\documentclass
[notitlepage,english,aps,floats,onecolumn,showpacs,nofootinbib,floatfix]
{revtex4-1}
\usepackage{pslatex}
\usepackage[T1]{fontenc}
\usepackage[latin1]{inputenc}
\usepackage{graphicx}
\usepackage{epsfig}
\usepackage{longtable}
\usepackage{float}
\usepackage{calc}
\usepackage{ifthen}

{
{
{
\newcommand{\bea}{\begin{eqnarray}}
\newcommand{\eea}{\end{eqnarray}}

\newcommand{\nc}{\newcommand}
\nc{\renc}{\renewcommand}
\nc{\eqs}[2]{\mbox{Eqs.~(\ref{#1},\,\ref{#2})}}
\nc{\eq}[1]{\mbox{Eq.~(\ref{#1})}}
\nc{\figs}[2]{\mbox{Figs.~(\ref{#1},\,\ref{#2})}}
\nc{\fig}[1]{\mbox{Fig~.(\ref{#1})}}
\nc{\be}[1]{\begin{equation} \mbox{$\label{#1}$}}
\nc{\ee}{\vspace{0.1cm}\end{equation}}

\newcommand{\bean}{\begin{eqnarray*}}
\newcommand{\eean}{\end{eqnarray*}}

\def\tdN{\tilde{N}}

\def\lae{\;^{<}_{\sim} \;} \def\gae{\; ^{>}_{\sim} \;}

\begin{document}
\title{Non-Minimally Coupled Inflation with Initial Conditions from a Pre-Inflation Anamorphic Contracting Era}
\author{John McDonald}
\email{j.mcdonald@lancaster.ac.uk}
\affiliation{Dept. of Physics, University of 
Lancaster, Lancaster LA1 4YB, UK}

\begin{abstract}

     Inflation due to a non-minimally coupled scalar field, as first proposed by Salopek, Bardeen and Bond (SBB), is in good agreement with the observed value of the spectral index and constraints on the tensor-to-scalar ratio. Here we explore the possibility that SBB inflation represents the late stage of a Universe which emerges from an early contracting era. We present a model in which the Universe smoothly transitions from an anamorphic contracting era to late-time SBB inflation without encountering a singular bounce. This corresponds to a continuous expansion in the Einstein frame throughout. We show that the anamorphic contracting era is able to provide the smooth superhorizon initial conditions necessary for subsequent SBB inflation to occur. The model predicts corrections to the non-minimal coupling, kinetic term and potential of SBB inflation which can observably increase the spectral index relative to its SBB prediction.

\end{abstract}
 
\maketitle

  Non-minimally coupled scalar field inflation models with a large non-minimal coupling and symmetric vacuum were first proposed by Salopek, Bardeen and Bond (SBB) in \cite{sbb} \footnote{A different class of non-minimally coupled inflation model was proposed earlier in \cite{spok} and \cite{igi}. Unlike the model of \cite{sbb}, the scalars in these models have large masses and expectation values in the present vacuum.}. The great advantage of such models is their ability use $\phi^4$ scalar potentials which have couplings of magnitude typical of particle physics models. This allows a conventional TeV-scale particle theory to account for inflation without the extremely small couplings encountered in minimally-coupled inflation models. For example, the Higgs boson \cite{hi} or a Higgs portal dark matter scalar \cite{lm,km} could account for inflation. In addition, SBB inflation predicts\footnote{Since SBB inflation is a single-field inflation model, it also predicts negligible non-Gaussianity, with $f_{NL} \sim  \eta \sim 10^{-2}$.} 
$n_{s} = 1 -2/\tdN - 3 / \tdN^2  = 0.966$  and $r =  12/\tdN^2  = 3.3\times 10^{-3}$ for $\tdN = 60$ (where $\tdN$ is the number of e-foldings in the Einstein frame), which is in very good agreement with the observed spectral index, $n_{s} = 0.9677 \pm 0.0060$ (68$\%$ CL, Planck TT + lowP + lensing), and is easily consistent with the upper bound on the tensor-to-scalar ratio,  $r_{0.002} < 0.11$ (95$\%$ CL, Planck TT + lowP + lensing) \cite{planckinf}. 

   If SBB inflation is indeed the correct model, it is natural to consider the origin of this era of observable inflation. Inflation requires a smooth, potential-dominated initial state over a region greater than the Hubble radius \cite{contractsm1}. One way this can be achieved is via an initial contracting era during which physical length scales contract less rapidly than $|H|^{-1}$, which requires a contracting era with an equation of state such that $\ddot{a} < 0$.  A stronger condition is that the contribution of the energy density to the Friedmann equation during contraction does not become dominated by the contribution of the Kasner-type metric anisotropies, which grows as $a^{-6}$ \cite{turok,brand}. The advantage of contraction is that it does not require an initial state that is smooth to begin with.

     Recently a model was proposed, the Anamorphic Universe \cite{au}, in which a contracting Universe smoothly transitions to an expanding Universe\footnote{Related ideas were previously explored in \cite{gasperini}.} at a finite value of $a$. This model allows a non-singular bounce which does not strongly amplify perturbations and anisotropies, in contrast to the case of a ghost condensate non-singular 
bounce \cite{xue}, and so can smooth the initial state of the expanding era on superhorizon scales. This model can be explicitly realized by a non-minimally coupled scalar model. In \cite
{au} was shown that, with appropriate choices for the conformal factor $\Omega(\phi)$, the scalar kinetic term $k(\phi)$ and the Jordan frame potential $V_{J}(\phi)$, it is possible to have an expansion in the Einstein frame which corresponds to a contraction in the physical Jordan frame (hence 'anamorphic'). In \cite{au} the objective was to produce a contracting Jordan frame model which is equivalent to inflation in the Einstein frame, in which case the contracting model can make predictions equivalent to a conventional inflation model. Our aim here is to use the anamorphic framework to  construct an      
initially contracting era that can evolve into late-time SBB inflation without encountering a singular bounce and which can create the initial conditions for SBB inflation. The transition from contraction to expansion in the Jordan frame corresponds to a smooth change in the form of expansion in the Einstein frame. 

  It is important to clearly distinguish betweeen the application of anamorphic contraction to inflation initial conditions and its use in the Anamorphic Universe model. In particular, the Anamorphic Universe is able to evade the creation of a multiverse via eternal inflation (since there is no inflation in the physical frame), whereas the present model reintroduces inflation at late times.

   We first make clear why Einstein frame expansion can correspond to Jordan frame contraction. This follows simply from the relation between the scale factor in the Einstein and Jordan frames, $a = \tilde{a}/\Omega $,  
where $\tilde{a}$ is the Einstein frame scale factor and $a$ is the Jordan frame scale factor.  
This follows from the definition of $\Omega$, $\tilde{g}_{\mu\nu} = \Omega^{2} g_{\mu\nu}$. Therefore if the Einstein frame scale factor is expanding from $\tilde{a}$ to $\tilde{a}_{c}$, the ratio of scale factors in the Jordan frame is 
\be{e1}  \frac{a_{c}}{a} = \frac{\Omega(a)}{\Omega(a_{c})} \times 
\frac{\tilde{a}_{c}}{\tilde{a}}    ~.\ee 
Therefore if the conformal factor $\Omega$ becomes larger more rapidly than the rate at which Einstein frame scale factor increases, the Jordan frame scale factor will contract i.e. the physical Universe will be a contracting Universe. We will use \eq{e1} as the basis of our analysis in the following. A more general method is given in \cite{au}.

   The model we consider here is an example of a model which interpolates between an anamorphic contracting era at large $\phi$ and the SBB inflation at small $\phi$.  In general, the scalar-tensor models of interest have the Jordan frame form (with $M_{Pl} = 1/(8 \pi G)^{1/2} = 1$)
\be{e2} S_{J} = \int d^4 x \sqrt{-g} \left[\frac{\Omega^2 R}{2} - \frac{k(\phi)}{2} \partial_{\mu}\phi \partial^{\mu} \phi - V_{J}(\phi) \right]  ~,\ee
where we follow \cite{au} in using the signature $(-,+,+,+)$. 
For the SBB model \cite{sbb}, valid at small $\phi$, 
\be{e3} \Omega^2 = 1 + \xi \phi^2 \;\;\;;\;\;\; 
k(\phi) = 1\;\;\;;\;\;\;  V_{J}(\phi) = \frac{\lambda \phi^4}{4}   ~,\ee 
while for the anamorphic contraction model, valid at large $\phi$,  we will consider
\be{e4} \Omega^2 = \alpha e^{-2A \phi} \;\;\;;\;\;\; 
k(\phi) = -\eta e^{-2A\phi} \;\;\;;\;\;\;  V_{J}(\phi) = \beta e^{-B\phi}   ~.\ee
Here $A$ and $B$ are positive. This differs from the simplest model of \cite{au} in that $(A,B) \rightarrow (-A,-B)$. This is necessary in order that $\phi$ is decreasing with time and so can transition to the SBB model at small $\phi$.

To smoothly transition between these limits, we will consider the following model ('interpolation model') which interpolates between the anamorphic contraction and SBB eras, 
\be{e5} \Omega^2 = 1 + \frac{\xi \phi^2}{\left(1 + \gamma \phi^{2} e^{2A\phi}\right)} \;\;\;;\;\;\; 
k(\phi) = \frac{1 - \gamma \phi^2}{1 + \gamma \phi^{2} e^{2A\phi}} \;\;\;;\;\;\;  V_{J}(\phi) = \frac{\lambda \phi^4}{4\left(1 + \gamma \phi^{2} e^{B\phi/2}\right)^2 }   ~.\ee 
In the Einstein frame this will correspond to a continuously expanding model. Therefore there is no amplification of perturbations or anisotropies calculated in the Einstein frame as the Universe transitions from contraction to expansion in the Jordan frame, in contrast to the case of a non-singular bounce due to ghost condensation. When $\phi < \phi_{c} = 1/\sqrt{\gamma}$, assuming that $A \phi_{c}$ and $B \phi_{c}$ are small compared to 1, \eq{e5} reduces to the SBB model, while at $\phi > \phi_{c}$ it becomes the anamorphic contraction model with a particular set of coefficients,  
\be{e6} \alpha = \frac{\xi}{\gamma} \;\;\;;\;\;\; \eta = 1 \;\;\;;\;\;\; \beta = \frac{\lambda}{4 \gamma^2}   ~.\ee   
During SBB inflation, $\phi_{\tilde{N}} = \sqrt{4 \tilde{N}/3 \xi}$. To have the correct magnitude of density perturbations,  we require that  $\xi \approx 10^{5} \lambda^{1/2}$, where we are most interested in the case where $\lambda$ can be large, $\lambda \sim 1$. Therefore, as long as $\sqrt{\gamma} << \sqrt{\xi/\tilde{N}}$, $\phi_{c} \gg \phi_{\tilde{N}}$ will be satisfied at $\tilde{N} \approx 60$ and the corrections to SBB inflation will be small. We will return these corrections later.   

In general, the Einstein frame action is
\be{e7}  S_{E} = \int d^4 x 
\sqrt{-\tilde{g}} 
\left[
\frac{\tilde{R}}{2}  
-\frac{3}{4 \Omega^4} 
\partial_{\mu} \Omega^2 \partial^{\mu} \Omega^2 
- \frac{k(\phi)}{2 \Omega^2} 
\partial_{\mu}\phi \partial^{\mu} \phi - 
\frac{V_{J}(\phi)}{\Omega^4} 
\right]  ~.\ee
In the case of the anamorphic contraction model the action becomes 
\be{e8}  S_{E} = \int d^4 x 
\sqrt{-\tilde{g}} 
\left[
\frac{\tilde{R}}{2}  
- \frac{1}{2}\left(6A^{2} - \frac{\eta}\alpha
\right)
\partial_{\mu}\phi \partial^{\mu} \phi 
- \frac{\beta e^{(4A-B) \phi}}{\alpha^2} 
\right]  ~.\ee
Provided that $6A^{2} > \eta/\alpha$, this has the correct sign of kinetic term in the Einstein frame, despite apparently being the wrong sign in the Jordan frame. Therefore, since we quantize in the Einstein frame, there is no problem of instability due to a ghost field. Rescaling $\phi$ to a canonically normalized scalar $\chi = 
(6A^{2} - \eta/\alpha)^{1/2} \phi$, the Einstein frame potential becomes 
\be{e9} V_{E} = \frac{\beta}{\alpha^2} \exp \left(\frac{(4A - B) \chi}{ (6A^{2} - \eta/\alpha)^{1/2} } \right)
~.\ee 
Since we want $\chi$ (and so $\phi$) to decrease with time in order to transition to SBB inflation at small $\phi$, we require that $4A > B$.
In order to have a model with an analytic solution, we will restrict attention to the case where $\chi$ is slow-rolling during the contracting phase. This is not essential but simply convenient. In this case the number of e-folds in the Einstein frame on rolling from $\chi$ to $\chi_{c}$ (where $\chi > \chi_{c}$) is
\be{e10}  \tilde{N} = -\int_{\chi}^{\chi_{c}} \frac{V_{E}}{V_{E}'} \, d \chi = \frac{\left(6A^{2} - \eta/\alpha\right)^{1/2}}{(4A - B)} \left(\chi - \chi_{c}\right)    ~.\ee
The ratio of conformal factors is 
\be{e12} \frac{\Omega}{\Omega_{c}} =  \exp\left(-\frac{A(\chi - \chi_{c}) }{\left(6A^2 - \eta/\alpha\right)^{1/2}} \right) ~.\ee
Thus the ratio of scale factors in the Jordan frame is
\be{e13} \frac{a_{c}}{a} = \frac{\Omega}{\Omega_{c}} e^{\tilde{N}} = 
\exp\left(-\Delta (\chi - \chi_{c}) \right) 
~,\ee
where
\be{e13a} \Delta = \frac{A}{\left(6 A^2 - \frac{\eta}{\alpha} \right)^{1/2} } -  \frac{\left(6 A^2 - \frac{\eta}{\alpha} \right)^{1/2} }{\left(4A - B\right)}   ~.\ee
Therefore the condition for contraction in the Jordan frame as the field rolls from $\chi$ to $\chi_{c}$ is $\Delta > 0$, which requires that
\be{e15} \frac{\eta}{\alpha} > 2A^2 + AB   ~.\ee
This is consistent with the result for the corresponding model in \cite{au}. Since $A$ and $B$ are positive, \eq{e15} can only be satisfied if the $\phi$ kinetic term in the Jordan frame has the wrong sign, $\eta > 0$. 
The condition for slow-rolling to be valid in the Einstein frame, $\tilde{\eta} \equiv |V_{E}''/V_{E}| < 1$, is satisfied if 
\be{e16}   (4A - B)^2 < 6A^{2} - \frac{\eta}{\alpha}    ~.\ee
Thus both conditions can be satisfied if $4A > B > 3A$ and $6 A^2 \approx \eta/\alpha$. 
In the interpolation model $\eta = 1$ and $\alpha = \xi/\gamma$. In this case both conditions can be satisfied if 
\be{e17} A \approx \frac{1}{\sqrt{6}} \left( \frac{\gamma}{\xi} \right)^{1/2}    ~.\ee 
Since it is assumed that $\gamma \ll \xi$ in order that the model tends to SBB inflation at small $\phi$, it follows that $A < 1$. We also assumed that $2 A \phi < 1$ and $B \phi/2 < 1$ at $\phi_{c} = 1/\sqrt{\gamma}$. From \eq{e17} we find that these are satisfied if $\xi > 4/3$. Since $\xi \sim 10^5$ in SBB inflation, this is easily satisfied. 
Thus there is a consistent slow-roll solution of the interpolation model in which the Universe undergoes anamorphic contraction in the Jordan frame at early times when $\phi < \phi_{c}$ and smoothly transitions to SBB inflation once 
$\phi > \phi_{c}$.

   We next check the condition for the early anamorphic contraction to produce the smooth initial conditions for SBB inflation on superhorizon scales. The strongest requirement is that the contribution of the anamorphic era energy density to the Friedmann equation in the Jordan frame grows more rapidly than the contribution of Kasner-type anisotropies during the contraction \cite{turok,brand}.  The contribution of the potential energy density to the Friedmann equation is proportional to $V_{J}/M_{Pl\; eff}^{2} = V_{J}/\Omega^{2}$, while the anisotropy contribution is proportional to $a^{-6}$ \cite{colley}. The potential contribution during anamorphic contraction evolves as 
\be{s1}  \frac{V_{J}}{\Omega^2} = \frac{\beta}{\alpha} \exp\left( -\frac{(B-2A) \chi}{\left(6 A^{2} - \frac{\eta}{\alpha} \right)^{1/2} }  \right)   
~.\ee
 From \eq{e13} it follows that 
\be{s2}   \frac{V_{J}}{\Omega^2} \propto a^{-r} \;\;\;;\;\;\; r =  \frac{(B-2A)}{\Delta \left(6 A^{2} - \frac{\eta}{\alpha} \right)^{1/2} }    ~.\ee
Requiring that $V_{J}/\Omega^2$ increases more rapidly that $a^{-6}$ during contraction then requires that 
\be{s3} (B-2A)(4A - B) > 6\left( A\left(4A-B\right) - \left(6 A^{2} - \frac{\eta}{\alpha} \right) \right)   ~.\ee 
From the slow-roll condition, \eq{e16}, the right side of \eq{s3} has an upper bound given by $6(4A-B)(B - 3A)$. Therefore the smoothing condition \eq{s3} is satisfied if
\be{s4}  \frac{16}{5} A > B   ~.\ee 
This is consistent with the range $4A > B > 3A$ for which contraction and slow-roll can both occur. We have also confirmed, although we do not show it here,  that the weaker smoothing condition, $\ddot{a} < 0$, is generally satisfied if $4A - B > 0$.

    We next consider the possible modification of the SBB inflation model due to the early contracting era. At $\phi \ll   \phi_{c}$, the functions in \eq{e5} become
\be{e18} \Omega^2(\phi) \approx 1 + \xi \phi^2 - \xi \gamma \;\phi^4 + \xi \gamma^2 \; \phi^6 \;\;;\;\; k(\phi) \approx 1 - 2 \gamma \;\phi^2  +2 \gamma^2 \; \phi^4 \;\;;\;\; V_{J} \approx \frac{\lambda}{4} \; \phi^4 -  \frac{\lambda\gamma}{2} \; \phi^6  +   
\frac{3 \lambda \gamma^2}{4} \phi^8
~,\ee
where we have included terms to next-to-leading order in $\gamma \phi^2$, as the leading-order terms cancel in the Einstein frame potential. (We have also assumed that $A \phi$ and $B \phi$ are small enough that $e^{2 A \phi}$ and $e^{B \phi/2}$ can be set equal to 1 in \eq{e5} when deriving \eq{e18}. In the Appendix this is shown to be true for the case of interest where the spectral index modification is large enough to be observable.) In the limit $\xi \gg \gamma$ we find that $n_{s}$ is given by (Appendix)
\be{e19} n_{s} = 1 - \frac{2}{\tilde{N}} + \frac{1}{3} \left(\frac{32}{3} \frac{\gamma \tilde{N}}{\xi}\right)^2       ~,\ee
where we have included the leading order correction to $n_{s}$.
Therefore the spectral index is increased relative to the SBB model. 
This imposes a significant constraint on the model. In order that the correction to $n_{s}$ is not larger than O(0.01), $\gamma$ must satisfy
\be{e20}  \gamma \; \lae \; 27 \left(\frac{60}{\tilde{N}}\right) \left(\frac{\xi}{10^5}\right)  ~.\ee
(The condition \eq{e17} then requires that $A \sim B \lae  10^{-2}$.) The critical value of $\phi$ at which the transition from contraction to expansion occurs, $\phi_{c} = 1/\sqrt{\gamma}$, therefore satisfies 
\be{e21} \phi_{c} \gae 0.2  \left(\frac{\tilde{N}}{60}\right)^{1/2} \left(\frac{10^5}{\xi}\right)^{1/2}  ~.\ee 
Thus $\phi_{c}$ is generally close to or larger than the Planck scale. In particular, if the transition from contraction to expansion occurs when $\phi$ is close to the Planck scale, corresponding to approximate equality in \eq{e21}, then the corrections to the spectral index can be large enough to be observable
 \footnote{In the case where the couplings of $\phi$ to Standard Model particles have dimensionally natural magnitudes $\sim \, 0.01-1$, the reheating temperature is well-defined \cite{hbb}. This means that $\tilde{N}$ can be tightly constrained, which makes it possible to accurately predict the spectral index in this model.}.

   In conclusion, we have shown that it is possible for non-minimally coupled SBB inflation to consistently emerge from an anamorphic contracting era. The model has a non-singular transition to expansion which can provide the smooth initial conditions necessary for SBB inflation. The model predicts modifications to the SBB model that can be large enough to produce an observable deviation of the spectral index from its SBB prediction. 

\section*{Acknowledgement} 
Partly supported by STFC via the Lancaster-Manchester-Sheffield Consortium for Fundamental Physics under STFC grant ST/J000418/1.

\section*{Appendix: The Spectral Index of the Interpolating Model} 
\renewcommand{\theequation}{A-\arabic{equation}}
 \setcounter{equation}{0}

Here we outline the calculation of the spectral index of the interpolating model, \eq{e19}. The Einstein frame potential from \eq{e18}, to leading order in $\gamma \phi^2$, is\footnote{There is a cancellation of the $O(\gamma \phi^2)$ contributions from $V_{J}$ and $\Omega^4$ to $V_{E}$.} 
\be{aa1} V_{E} \approx \frac{V_{J_{\,0}}}{\Omega_{0}^4} \left(1 + 4 \gamma^2 \phi^4 \right)  ~,\ee
where $V_{J_{\,0}} = \lambda \phi^{4}/4$ and $\Omega_{0}^{2} = \left(1 + \xi \phi^2\right)$. The $\phi$ field must be transformed to the canonically normalized $\chi$ field in order to compute the slow-roll parameters. The $\phi$ kinetic term in the Einstein frame is 
\be{aa2} -\frac{1}{2} \left[ \frac{3}{2 \Omega^4} \left( \frac{\partial \Omega^2}{\partial \phi} \right)^2 + \frac{k(\phi)}{\Omega^2} \right] \partial_{\mu}\phi \partial^{\mu}\phi   ~.\ee
To leading-order in $\gamma \phi^2$, where $\gamma \phi^2 \ll 1$, this becomes 
\be{aa3} 
-\frac{1}{2} \left[ \frac{6}{\phi^2} \left(1 - 2 \gamma \phi^2\right) \right] \partial_{\mu}\phi \partial^{\mu}\phi ~.\ee
Thus the canonically normalized field $\chi$ is related to $\phi$ by 
\be{aa4}  \frac{\partial \chi}{\partial \phi} \approx  \frac{\sqrt{6}}{\phi}\left(1 - \gamma \phi^2\right)   ~.\ee
The derivatives of $V_{E}$ with respect to $\chi$, which determine the slow-roll parameters, are then
\be{aa5} \frac{\partial V_{E}}{\partial \chi} = \frac{\partial \phi}{\partial \chi} \frac{\partial V_{E}}{\partial \phi}  
\approx \frac{\lambda}{4 \sqrt{6}\, \xi^2} \left( \frac{4}{\xi \phi^2} + 16 \gamma^2 \phi^4 + \frac{4 \gamma}{\xi} \right)
~\ee
and 
\be{aa6} \frac{\partial^{2} V_{E}}{\partial \chi^{2}} = 
\frac{\partial \phi}{\partial \chi} \frac{\partial}{\partial \phi}\left(  \frac{\partial V_{E}}{\partial \chi} \right) \approx 
\frac{\lambda}{24 \xi^2} \left( - \frac{8}{\xi \phi^2} + 64 \gamma^2 \phi^4 - \frac{8 \gamma}{\xi} \right)  ~.\ee 
Thus 
\be{aa7} \eta = \frac{V_{E}^{''}}{V_{E}} \approx  
\left(-\frac{4}{3 \xi \phi^2} + + \frac{32}{3} \gamma^2 \phi^4 - \frac{4}{3} \frac{\gamma}{\xi} \right) (1 - 4 \gamma^2 \phi^4)   ~.\ee 
Using the relation\footnote{This relation is not significantly modified by the $\gamma \phi^2$ corrections in the case of interest.} between $\phi$ and $\tilde{N}$, $\phi = \sqrt{4 \tilde{N}/3 \xi}$, this becomes
\be{aa8} \eta \approx \left(-\frac{1}{\tilde{N}} + \frac{\left(32 \times 16\right)}{27} \frac{\gamma^2 \tilde{N}^2}{\xi^2} - \frac{4 \gamma}{3 \xi} \right) \left(1 - \frac{64}{9}  \frac{\gamma^2 \tilde{N}^2}{\xi^2} \right)    ~\ee 
Thus, to leading-order in $1/\tilde{N}$ and $\gamma \phi^2$, 
\be{aa9}  \eta \approx -\frac{1}{\tilde{N}} + \frac{1}{6} \left(\frac{32}{3}\right)^2  \left( \frac{\gamma \tilde{N}}{\xi} \right)^2   ~.\ee
The $\epsilon$ contribution to the spectral index is $O(1/\tilde{N}^2)$ and so is negligible. Thus, to leading-order, the spectral index is given by
\be{aa10} n_{s} \approx 1 + 2 \eta \approx 1 - \frac{2}{\tilde{N}} + \frac{1}{3} \left(\frac{32}{3}\right)^2  \left( \frac{\gamma \tilde{N}}{\xi} \right)^2 ~.\ee

  We finally show that the $e^{2 A \phi}$ and $e^{B \phi/2}$ terms in \eq{e5}  can be set equal to 1 when deriving \eq{e18}. Assuming that $A\phi \ll 1 $ and $B \phi \ll 1$ (this is easily satisfied since $\phi = \sqrt{4 \tilde{N}/3 \xi} \ll 1$ and $A \sim B \lae 10^{-2}$), the full form of \eq{e21}, to leading- order in $A \phi$ and $B \phi$, is   
\be{aa11}  \Omega^2(\phi) \approx 1 + \xi \phi^2 - \xi \gamma \;\phi^4 - 2 A \gamma \xi \; \phi^5 + \xi \gamma^2 \; \phi^6 \;\;;\;\; k(\phi) \approx 1 - 2 \gamma \;\phi^2 - 2 A \gamma \; \phi^3 +2 \gamma^2 \; \phi^4 \;\;;\;\; V_{J} \approx \frac{\lambda}{4} \; \phi^4 -  \frac{\lambda\gamma}{2} \; \phi^6 
- \frac{\lambda \gamma B}{4} \phi^7  +   
\frac{3 \lambda \gamma^2}{4} \phi^8
~. \ee
Since $A \approx B$, the general condition for the $A$ and $B$ correction terms to be negligible is $A \ll \gamma \phi$. For the case where the $n_{s}$ correction term is large enough to be observable, corresponding to equality in   
 \eq{e20}, we have $\gamma \sim 10$. Using $\phi = \sqrt{4 \tilde{N}/3 \xi}\,$, and with $A$ given by \eq{e17}, the condition for the $A$ and $B$ corrections to be negligible becomes
\be{aa12} \gamma \gg \frac{1}{\tilde{N}}  ~.\ee
This is easily satisfied when $\gamma \sim 10$ and $\tilde{N} \approx 60$.

\end{document}